\documentclass{aastex}
\usepackage{emulateapj5}
\usepackage{apjfonts}
\usepackage{epsf}
\slugcomment{UTAP-481}
\shorttitle{STATISTICAL SIGNIFICANCE OF SMALL SCALE ANISOTROPY}
\shortauthors{YOSHIGUCHI ET AL.}
\begin{document}
%%%%%%%%%%%%%%%%%%%%%%%%%%%%%%%%%%%%%%%%%%%%%%%%%%%%%%%%%%%%%%%%%%%%%%
%%%%%%%%%%%%%%%%%%%%%%%%%%%%%%%%%%%%%%%%%%%%%%%%%%%%%%%%%%%%%%%%%%%%%%
\title{Statistical Significance of Small Scale Anisotropy in Arrival
Directions of Ultra-High Energy Cosmic Rays}
%%%%%%%%%%%%%%%%%%%%%%%%%%%%%%%%%%%%%%%%%%%%%%%%%%%%%%%%%%%%%%%%%%%%%%
%%%%%%%%%%%%%%%%%%%%%%%%%%%%%%%%%%%%%%%%%%%%%%%%%%%%%%%%%%%%%%%%%%%%%%
%
%%%%%%%%%%%%%%%%%%%%%%%%%%%%%%%%%%%%%%%%%%%%%%%%%%%%%%%%%%%%%%%%%%%%%%
\author{Hiroyuki Yoshiguchi\altaffilmark{1}, Shigehiro
Nagataki\altaffilmark{2}, and Katsuhiko Sato\altaffilmark{1,3}}

\altaffiltext{1}{Department of Physics, School of Science, the University
of Tokyo, Tokyo 113-0033, Japan}
\altaffiltext{2}{Yukawa Institute for Theoretical Physics, Kyoto University, Kyoto 606-8502, Japan}
\altaffiltext{3}{Research Center for the Early Universe, School of
Science, the University of Tokyo, Tokyo 113-0033, Japan}

\email{hiroyuki@utap.phys.s.u-tokyo.ac.jp}
%%%%%%%%%%%%%%%%%%%%%%%%%%%%%%%%%%%%%%%%%%%%%%%%%%%%%%%%%%%%%%%%%%%%%%
%
\received{}
\accepted{}
\begin{abstract}
Recently, the High Resolution Fly's Eye (HiRes) experiment claims that
there is no small scale anisotropy in the arrival distribution of
ultra-high energy cosmic rays (UHECRs) above $E>10^{19}$ eV contrary
to the Akeno Giant Air Shower Array (AGASA) observation.
In this paper, we discuss the statistical significance of this
discrepancy between the two experiments.
We calculate arrival distribution of UHECRs above $10^{19}$ eV
predicted by the source models constructed using the Optical Redshift
Survey galaxy sample.
We apply the new method developed by us for
calculating arrival distribution in the presence of the
galactic magnetic field.
The great advantage of this method is that it enables us to calculate
UHECR arrival distribution with lower energy ($\sim 10^{19}$ eV) than
previous studies within reasonable time by following only the
trajectories of UHECRs actually reaching the earth.
It has been realized that the small scale anisotropy observed by the
AGASA can be explained with the source number density $\sim 10^{-5
\sim -6}$ Mpc$^{-3}$ assuming weak extragalactic magnetic field ($B
\le 1$ nG).
We find that the predicted small scale anisotropy for this source
number density is also consistent with the current HiRes data.
We thus conclude that the statement by the HiRes experiment that they
do not find small scale anisotropy in UHECR arrival distribution is
not statistically significant at present.
We also show future prospect of determining the source number density
with increasing amount of observed data.
\end{abstract} 
\keywords{cosmic rays --- methods: numerical --- ISM: magnetic fields ---
galaxies: general --- large-scale structure of universe}
%
%%%%%%%%%%%%%%%%%%%%%%%%%%%%%%%%%%%%%%%%%%%%%%%%%%%%%%%%%%%
%%%%%%%%%%%%%%%%%%%%%%%%%%%%%%%%%%%%%%%%%%%%%%%%%%%%%%%%%%%
\section{INTRODUCTION} \label{intro}
%%%%%%%%%%%%%%%%%%%%%%%%%%%%%%%%%%%%%%%%%%%%%%%%%%%%%%%%%%%
%%%%%%%%%%%%%%%%%%%%%%%%%%%%%%%%%%%%%%%%%%%%%%%%%%%%%%%%%%%

The small scale anisotropy in the observed arrival distribution of
ultra-high energy cosmic rays (UHECRs) is key in identifying their
sources which will tell us a great deal about acceleration
mechanisms, composition of UHECRs, and so on.
The Akeno Giant Air Shower Array (AGASA) observation reveals
the existence of the small scale clusterings in the isotropic arrival
distribution of UHECRs \citep{takeda99,takeda01}.
The current AGASA data set of 57 events above 4 $\times 10^{19}$ eV
contains four doublets and one triplet within
a separation angle of 2.5$^\circ$.
Chance probability to observe such clusters under an isotropic
distribution is only about 1 $\%$ \citep*{hayashida00,takeda01}.

On the other hand, the recent claim by the High Resolution Fly's Eye
\citep*[HiRes; ][]{wilkinson99} experiment is that there is no small
scale anisotropy in the UHECR arrival distribution above
$E>10^{19}$ eV (162 events) which is observed by a stereo air
fluorescence detector \citep{finley03}.
The discrepancy between the AGASA and HiRes involves a number of
consideration.
Among them are the different numbers of the events, the possibility of
the measured energies being shifted between the two experiments,
\citep*{marco03} and the different angular resolutions.
The purpose of this paper is to discuss the first possibility.
That is, we consider whether the discrepancy between the two
experiments is statistically significant or not at present.

There is furthermore the contradiction on the presence or absence of
the GZK cutoff \citep*{greisen66,zatsepin66} in the cosmic-ray energy
spectrum due to photopion production with the photons of the
cosmic microwave background (CMB).
The HiRes spectrum shows the GZK cutoff \citep*{abu02}, but the AGASA
spectrum does not \citep*{takeda98}.
The statistical significance of this discrepancy is already discussed
in \cite{marco03} and shown to be low at present.
This issue is left for future investigation by new large-aperture
detectors under development, such as South and North Auger project
\citep*{capelle98}, the EUSO \citep*{euso92}, and the OWL
\citep*{owl00} experiments.

When we consider the statistical significance of the small scale
anisotropy, we need to compare the observations with numerical
calculations predicted by some kind of source distributions.
In this paper, we use the Optical Redshift Survey \citep*[ORS;
][]{santiago95} galaxy sample to construct realistic source models of
UHECRs.
This sample has been adopted in our previous studies
\citep*{yoshiguchi03a,yoshiguchi03c,yoshiguchi03d}.
It has been realized that the small scale anisotropy well reflect the
number density of UHECR sources
\citep*{yoshiguchi03a,yoshiguchi03c,blasi04}.
And also, it is shown that the small scale anisotropy
above $E=4\times 10^{19}$ eV observed by the AGASA can be explained
with the source number density $\sim 10^{-5 \sim -6}$ assuming weak
extragalactic magnetic field ($B \le 1$ nG, EGMF) and that observed
UHE particles are protons
\citep*{yoshiguchi03a,yoshiguchi03c,blasi04}.
We thus take the source number density as a parameter of our source
model, and discuss the statistical significance of the small scale
anisotropy by considering to what extent we can determine the source
number density from the comparison of our model predictions with the
HiRes observation.

As mentioned above, the claim by the HiRes experiment on the small
scale anisotropy is based on the data of UHECR arrival directions
above $10^{19}$ eV.
UHE protons at such energies are significantly deflected by the galactic
magnetic field (GMF) by a few $- \sim 10$ degrees.
In order to accurately calculate the expected UHECR arrival
distribution and compare with the observations, this effect
should be taken into account.

In our recent work \citep*{yoshiguchi03d}, we presented a new method
for calculating UHECR arrival distribution which can be applied to
several source location scenarios including modifications by the GMF.
Brief explanation of our method is as follows.
We numerically calculate the propagation of anti-protons from the
earth toward the outside of the Galaxy (we set a sphere centered
around the Galactic center with radius $r_{\rm src}=$40 kpc as the
boundary condition), considering the deflections due to the GMF.
The anti-protons are ejected isotropically from the earth.
By this calculation, we can obtain the trajectories and the sky map of
position of anti-protons that have reached to the boundary at radius
$r_{\rm src}=$ 40 kpc.

Next, we regard the obtained trajectories as the ones of PROTONs from
the outside of the galaxy toward the earth.
Also, we regard the obtained sky map of position of anti-proton at the
boundary as relative probability distribution (per steradian) for
PROTONs to be able to reach to the earth for the case in which the
flux of the UHE protons from the extra-galactic region is isotropic
(in this study, this flux corresponds to the one at the boundary
$r_{\rm src}=40$ kpc).
The validity of this treatment is supported by the Liouville's
theorem.
When the flux of the UHE protons at the boundary is anisotropic (e.g.,
the source distribution is not isotropic), this effect can be included
by multiplying this effect (that is, by multiplying the probability
density of arrival direction of UHE protons from the extra-galactic
region at the boundary) to the obtained relative probability
density distribution mentioned above.

By adopting this new method, we can consider only the trajectories of
protons which arrive to the earth, which, of course, helps us to save
the CPU time efficiently and makes calculation of propagation of CRs
even with low energies ($\sim 10^{19}$ eV) possible within a
reasonable time.

With this method, we calculate the arrival distribution of UHE protons
above $10^{19}$ eV for our source models.
We also consider the energy loss processes when UHE protons propagate
in the intergalactic space.
Using the two point correlation function as statistical quantities for
the small scale anisotropy, we compare our model prediction with the
HiRes observation.
We then discuss to what extent we can determine the source number
density of UHECRs by the current HiRes data.
We also briefly discuss future prospect of determining the source
number density with the event number expected by future experiments.

In section~\ref{gmf}, we introduce the GMF model.
We explain the method for calculating UHECR arrival distribution in
section~\ref{method}.
Results are shown in section~\ref{result}.
In section~\ref{summary}, we summarize the main results.

%%%%%%%%%%%%%%%%%%%%%%%%%%%%%%%%%%%%%%%%%%%%%
\section{GALACTIC MAGNETIC FIELD} \label{gmf}
%%%%%%%%%%%%%%%%%%%%%%%%%%%%%%%%%%%%%%%%%%%%%
In this study, we adopt the GMF model used in
\cite{stanev02,yoshiguchi03d,yoshiguchi04},
which is composed of the spiral and the dipole field.
In the following, we briefly introduce this GMF model.

Faraday rotation measurements indicate that the GMF in the
disk of the Galaxy has a spiral structure with field reversals
at the optical Galactic arms \citep{beck01}.
We use a bisymmetric spiral field (BSS) model, which is favored
from recent work \citep{han99,han01}.
The Solar System is located at a distance
$r_{\vert\vert}=R_\oplus=8.5$ kpc from the center of the Galaxy in the
Galactic plane.
The local regular magnetic field in the vicinity of the Solar System
is assumed to be 
$B_{\rm Solar} \sim 1.5~\mu{\rm G}$ in the direction $l=90^{\rm o}+p$
where the pitch angle is $p=-10^{\rm o}$ \citep{han94}.

In the polar coordinates $(r_{\vert\vert},\phi)$,
the strength of the spiral field in the Galactic plane is given by
\begin{equation}
B(r_{\vert\vert},\phi)=
B_0~\left({R_\oplus \over r_{\vert\vert}}\right)~
\cos\left(\phi - \beta \ln {r_{\vert\vert}\over r_0} \right)
\end{equation}  
where $B_0=4.4~\mu$G, $r_0= 10.55$ kpc and $\beta=1/\tan p=-5.67$.
The field decreases with Galactocentric distance as $1/r_{\vert\vert}$ 
and it is zero for $r_{\vert\vert}>20$ kpc. 
In the region around the Galactic center ($r_{\vert\vert} < 4$ kpc) 
the field is highly uncertain, and thus assumed to be constant and
equal to its value at $r_{\vert\vert}=4$ kpc.

The spiral field strengths above and below the Galactic plane are taken to
decrease exponentially with two scale heights \citep{stanev96}
\begin{equation}
\vert B(r_{\vert\vert},\phi,z)\vert = 
\vert B(r_{\vert\vert},\phi)\vert 
\left\{
\begin{array}{lcl}
\exp(-z) :  & \vert z \vert \leq 0.5~ {\rm kpc} \\ 
\exp( \frac{-3}{8})~\exp(\frac{-z}{4}) :  & \vert z \vert > 0.5~ {\rm kpc} 
\end{array}
\right.
\label{eq:b-height}
\end{equation} 
where the factor $\exp(-3/8)$ makes the field continuous in $z$.
The BSS spiral field we use is of even parity, that is,
the field direction is preserved at disk crossing.

Observations show that 
the field in the Galactic halo is much weaker than that in the disk.
In this work we assume that the regular field corresponds to a 
A0 dipole field as suggested in \citep{han02}.
In spherical coordinates $(r,\theta,\varphi)$ 
the $(x,y,z)$ components of the halo field are given by:
\begin{eqnarray}
B_x=-3~\mu_{\rm G}~{\sin\theta \cos\theta \cos\varphi}/
r^3 \nonumber \\ 
B_y=-3~\mu_{\rm G}~{\sin\theta \cos\theta \sin\varphi}/
r^3 \\
B_z=\mu_{\rm G}~{(1-3\cos^2\theta)}/r^3 ~~~~~~~ \nonumber
\label{eq:bdipole}
\end{eqnarray}
where $\mu_{\rm G}\sim 184.2~{\rm \mu G~kpc^3}$ is the magnetic moment
of the Galactic dipole.
The dipole field is very strong in the central region 
of the Galaxy, but is only 0.3 $\mu$G in the vicinity
of the Solar system, directed toward the North Galactic Pole.

There may be a significant turbulent component, $B_{\rm ran}$,
of the Galactic magnetic field.
Its field strength is difficult to measure and results found in
literature are in the range of $B_{\rm ran} = 0.5 \dots 2 B_{\rm reg}$
\citep{beck01}.
However, we neglect the random field throughout the paper for
simplicity.
Possible dependence of the results on the assumption is discussed in
the final section.

%%%%%%%%%%%%%%%%%%%%%%%%%%%%%%%%%%%%%%%%%%%%%
\section{NUMERICAL METHOD} \label{method}
%%%%%%%%%%%%%%%%%%%%%%%%%%%%%%%%%%%%%%%%%%%%%

%-----------------------------------------------------------------
\subsection{Propagation of UHE protons in the Intergalactic Space}
\label{sim}
%-----------------------------------------------------------------

The energy spectrum of UHECRs injected at extragalactic sources is
modified by the energy loss processes when they propagate in the
intergalactic space. 
In this subsection, we explain the method of Monte Carlo simulations
for propagation of UHE protons in intergalactic space.

We assume that the composition of UHECRs are protons which are
injected with a power law spectrum within the range of ($10^{19}$ -
$10^{22}$)eV.
We inject 10000 protons in each of 31 energy bins (10 bins per decade
of energy).
Then, UHE protons are propagated including the energy loss processes
(explained below) over $3$ Gpc for $15$ Gyr.
We take a power law index as 2.6 to fit the predicted energy
spectrum to the one observed by the HiRes \citep{marco03}.

UHE protons below $\sim 8 \times 10^{19}$ eV lose their energies
mainly by pair creations and adiabatic energy losses, and above it by
photopion production \citep*{berezinsky88,yoshida93} in collisions
with photons of the CMB.
We treat the adiabatic loss as a continuous loss process.
We calculate the redshift $z$ of source at a given distance using the
cosmological parameters $H_0=71$ km s$^{-1}$ Mpc$^{-1}$,
$\Omega_m=0.27$, and $\Omega_{\Lambda}=0.73$.
Similarly, the pair production can be treated as a continuous loss process
considering its small inelasticity ($\sim 10^{-3}$).
We adopt the analytical fit functions given by \cite{chodorowski92}
to calculate the energy loss rate for the pair production
on isotropic photons.

On the other hand, protons lose a large fraction of their energy in
the photopion production.
For this reason, its treatment is very important.
We use the interaction length and the energy distribution of
final protons as a function of initial proton energy
which is calculated by simulating the photopion
production with the event generator SOPHIA \citep*{sophia00}.
The same approach has been adopted in our previous studies
\citep*{ide01,yoshiguchi03a,yoshiguchi03c,yoshiguchi03d}.

In this study, we neglect the effect of the EGMF because of the
following two reasons.
First, numerical simulations of UHECR propagation in the EGMF
including lower energy ($\sim 10^{19}$ eV) ones take a long CPU time.
Secondly, we show in our previous study that small scale clustering
observed by the AGASA can be well reproduced in the case of weak EGMF
$(B < 1{\rm nG})$ \citep*{yoshiguchi03a,yoshiguchi03c}.
The purpose of this paper is to discuss whether the claim by the HiRes
experiment is contrary to the AGASA result with statistical
significance.
We should calculate the arrival distribution of UHECRs in the
situation that the AGASA experiment can be explained.

Of course, effects of strong EGMF (say, $\sim 100$ nG) on UHECR
propagation are intensively investigated
\citep*{sigl03,sigl03b,sigl04,aloisio04}, and can be hardly
excluded considering the limited amount of observational data on the
EGMF.
However, we assume extremely weak EGMF throughout the paper.

%-----------------------------------------------------------------
\subsection{Source Distribution}
\label{method_source}
%-----------------------------------------------------------------

In this study, we apply the method for calculating the UHECR arrival
distribution with modifications by the GMF
(section~\ref{calc_arrival}) to our source location scenario, which is
constructed by using the ORS \citep*{santiago95} galaxy catalog.

In order to calculate the distribution of arrival
directions of UHECRs realistically, there are two key elements
of the galaxy sample to be corrected.
First, galaxies in a given magnitude-limited sample are biased tracers
of matter distribution because of the flux limit \citep*{yoshiguchi03b}.
Although the sample of galaxies more luminous than $-20.5$ mag is
complete within 80 $h^{-1}$ Mpc (where $h$ is the Hubble constant
divided by 100 km s$^{-1}$ and we use $h=0.71$), it does not contain
galaxies outside it for the reason of the selection effect.
We distribute sources of UHECRs outside 80 $h^{-1}$ Mpc homogeneously.
Their number density is set to be equal to that inside 80 $h^{-1}$
Mpc.

Secondly, our ORS sample does not include galaxies
in the zone of avoidance ($|b|<20^{\circ}$).
In the same way, we distribute UHECR sources in this region homogeneously,
and calculate its number density from the number of galaxies in the
observed region.

As mentioned above, we take the number density of UHECR sources as our
model parameter when we construct the source models.
For a given number density, we randomly select galaxies from the above
sample with probability proportional to the absolute luminosity of
each galaxy.
In this paper, the model parameters are taken to be $10^{-3}$,
$10^{-4}$, $10^{-5}$, and $10^{-6}$ Mpc$^{-3}$.

%-----------------------------------------------------------------
\subsection{Calculation of the UHECR Arrival Distribution with
modifications by the GMF}
\label{calc_arrival}
%-----------------------------------------------------------------

This subsection provides the method of calculation of UHECR
arrival distribution with modifications by the GMF.
The explanations largely follows our recent paper
\citep*{yoshiguchi03d}.
We start by injecting anti-protons from the earth isotropically, and
follow each trajectory until

1. anti-proton reaches a sphere of radius $40$ kpc centered at the galactic
   center, or

2. the total path length traveled by anti-proton is larger than $200$
   kpc,

by integrating the equations of motion in the magnetic field.
It is noted that we regard these anti-protons as PROTONs injected from
the outside of the Galaxy toward the earth.
The number of propagated anti-proton is 2,000,000.
We have checked that the number of trajectories which are stopped by
the limit (2) is smaller than $0.1$\% of the total number.
The energy loss of protons can be neglected for these distances.
Accordingly, we inject anti-protons with injection spectrum similar to
the observed one $\sim E^{-2.7}$.
(Note that this is not the energy spectrum injected at extragalactic
sources.)

\begin{figure*}
\begin{center}
%\leavevmode
\epsscale{2.0}
\plotone{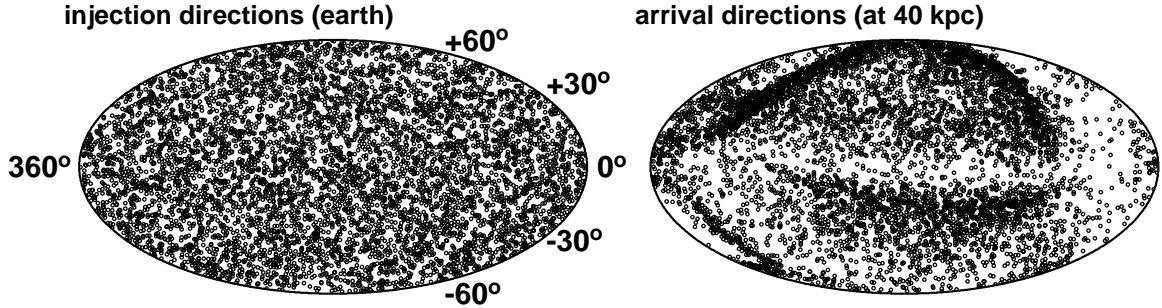} 
\caption{
Arrival directions of anti-protons with $E>10^{19.0}$
eV at the sphere of Galactocentric radius of $40$ kpc (right panel)
in the galactic coordinate.
The anti-protons are injected at the earth isotropically (as shown in
the left panel) with an injection spectrum $E^{-2.7}$.
\label{event}}
\end{center}
\end{figure*}

The result of the velocity directions of anti-protons at the sphere of
radius $40$ kpc is shown in the right panel of figure~\ref{event} in
the galactic coordinate.
From Liouville's theorem, if the cosmic-ray flux outside the Galaxy is
isotropic, one expects an isotropic flux at the earth even in the
presence of the GMF.
This theorem is confirmed by numerical calculations shown in figure 6
of \cite{stanev02}, which is the same figure as our figure~\ref{event}
except for threshold energy.
Thus, the mapping of the velocity directions in the right panel of
figure~\ref{event} corresponds to the sources which actually give rise
to the flux at the earth in the case that the sources (including ones
which do not actually give rise to the flux at the earth) are
distributed uniformly.

We calculate the UHECR arrival distribution for our source scenario
using the numerical data of the propagation of UHE anti-protons in the
Galaxy.
Detailed method is as follows.
At first, we divide the sky into a number of bins with the same solid
angle.
The number of bins is taken to be $360 (l) \times 200 (b)$.
We then distribute all the trajectories into each bin according to
their directions of velocities (source directions) at the sphere of
radius $40$ kpc.
Finally, we randomly select trajectories from each bin with
probability $P_{\rm selec} (j,k,E)$ defined as
\begin{equation}
P_{\rm selec} (j,k,E) \propto \sum _{i} \frac{1}{d_i^2} \, 
\frac{dN/dE(d_i,E)}{E^{-2.7}}.
\label{def_pselec} 
\end{equation}
Here subscripts j and k distinguish each cell of the sky, $d_i$ is
distance of each galaxy within the cell of (j,k), and the summation
runs over all of the galaxies within it.
$E$ is the energy of proton, and $dN/dE(d_i,E)$ is the energy
spectrum of protons at our galaxy injected at a source of distance
$d_i$.

The normalization of $P_{\rm selec} (j,k,E)$ is determined so as to
set the total number of events equal to a given number, for example,
the event number of the current HiRes data.
When $P_{\rm selec}>1$, we newly generate events with number of
$(P_{\rm selec}-1) \times N(j,k)$, where $N(j,k)$ is the number of
trajectories within the sky cell of $(j,k)$.
The arrival angle of newly generated proton (equivalently, injection
angle of anti-proton) at the earth is calculated by adding a normally
distributed deviate with zero mean and variance equal to the
experimental resolution $2.8^{\circ}$ $(1.8^{\circ})$ for $E>10^{19}$
eV $(4 \times 10^{19} {\rm eV})$ to the original arrival angle.

%-------------------------------------------------------------------------
\subsection{Statistical Methods}\label{statistics}
%-------------------------------------------------------------------------

A standard tool for searching the small scale anisotropy is the two
point correlation function.
This subsection provides the explanation of this function.

We start from a set of simulated events.
For each event, we divide the sphere into concentric bins of
angular size $\Delta \theta$, and count the number of events falling
into each bin.
We then divide it by the solid angle of the corresponding bin,
that is,
\begin{eqnarray}
N ( \theta ) = \frac{1}{2 \pi | \cos \theta  - \cos (\theta + \Delta \theta)
|} \sum_{ \theta
\le  \phi \le \theta + \Delta \theta }  1 \;\;\; [ \rm  sr ^{-1} ],
\label{eqn100}
\end{eqnarray}
where $\phi$ denotes the separation angle of the two events.
$\Delta \theta$ is taken to be $2^{\circ}$ in this analysis.
We also evaluate $N_{\rm uni} (\theta)$ for an equal number of events
simulated in a Monte Carlo for uniform source distribution.
The estimator for the correlation function is then defined as
$w(\theta)=N (\theta)/N_{\rm uni} (\theta) -1$.

The two point correlation function of the HiRes data with 164 events
above $10^{19}$ eV do not show significant small scale anisotropy, and
is consistent with isotropic source distribution within $1 \sigma$
confidence level.
We thus compare the two point correlation functions predicted by our
source models with that expected for isotropic source distribution
rather than the HiRes data itself.
When we consider to what extent we can determine the source number
density of UHECRs, we have to quantify deviations of predictions
by our source models from isotropic sources.
To do so, we define the variable $\chi^2$ as
\begin{eqnarray}
\chi^2 = \frac{1}{N_{\rm bin}} \sum_{i=1}^{i=N_{\rm bin}} \frac{(w_{\rm
th} ( \theta_{\rm i} ) - w_{\rm uni} ( \theta_i ))^2}{(\sigma_{\rm
i}^{\rm th})^2 + (\sigma_{\rm i}^{\rm uni})^2},
\label{chi}
\end{eqnarray}
where  $\theta_{\rm i}$ is the angle at i-th bin.
$\sigma_{\rm i}^{\rm th}$ and $\sigma_{\rm i}^{\rm uni}$ are the
statistical fluctuations of $w_{\rm th}$ and $w_{\rm uni}$ at the
angle $\theta_{\rm i}$.
The summation is taken over the angular bins, whose number is denoted
by $N_{\rm bin}$.

Even if we specify the source number density, source distribution
itself can not be determined because of randomness when we select
galaxies from the ORS sample.
Therefore, we first evaluate $\chi^2$ for a source distribution in the
case of a given number density of UHECR sources.
We then repeat such calculations for a number of realizations of the
source selection, and obtain the probability distribution of $\chi^2$
for a source number density.
The source selections are performed 100 times for all the source
number densities.

\section{RESULTS} \label{result}
%%%%%%%%%%%%%%%%%%%%%%%%%%%%%%%%%%%%%%%%%

%-------------------------------------------------------------------------
\subsection{Arrival Distribution of UHECRs}
\label{arrival}
%-------------------------------------------------------------------------

\begin{figure*}
\begin{center}
%\leavevmode
\epsscale{2.0}
\plotone{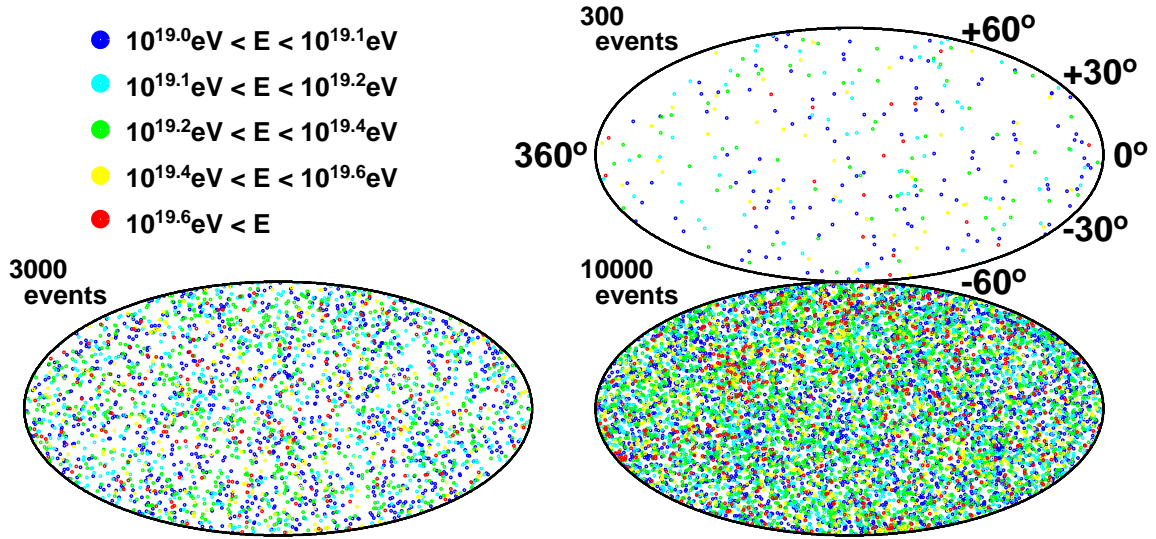} 
\caption{
Arrival directions of protons with $E>10^{19}$ eV predicted by a
source distribution for the source number density $10^{-3}$
Mpc$^{-3}$ in Galactic coordinates.
Events are shown by color according to their energies.
\label{image_s3}}
\end{center}
\end{figure*}

\begin{figure*}
\begin{center}
%\leavevmode
\epsscale{2.0}
\plotone{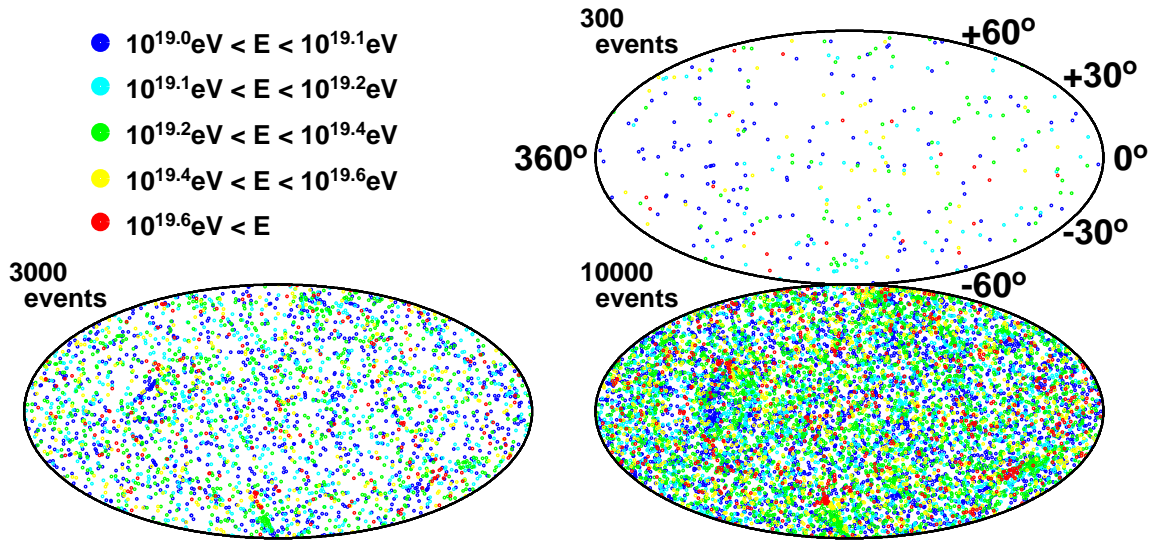} 
\caption{
Same as figure \ref{image_s3}, but for the source number density
$10^{-5}$ Mpc$^{-3}$.
\label{image_s5}}
\end{center}
\end{figure*}

In this subsection, we present the results of arrival distribution of
UHECRs above $10^{19}$ eV obtained by using the method explained in
section \ref{calc_arrival}.
Figure \ref{image_s3} and \ref{image_s5} show realizations of the
event generations in the case of the source number density $10^{-3}$
and $10^{-5}$ Mpc$^{-3}$, respectively.
The events are shown by color according to their energies.
Note that the event number of HiRes data is 162 ($E>10^{19}$ eV).
This roughly corresponds to 300 events of figure \ref{image_s3} and
\ref{image_s5}, considering the difference of the range $\delta$
(declination) between the observation and numerical calculation.

%\begin{figure*}
%\begin{center}
%\leavevmode
%\epsscale{2.0}
%\plotone{image_s4.eps} 
%\caption{
%\label{image_s4}}
%\end{center}
%\end{figure*}

%\begin{figure*}
%\begin{center}
%\leavevmode
%\epsscale{2.0}
%\plotone{image_s6.eps} 
%\caption{
%\label{image_s6}}
%\end{center}
%\end{figure*}

As is evident from figure \ref{image_s3} and \ref{image_s5}, arrival
distributions of UHECRs are quite isotropic.
The harmonic analysis to the right ascension distribution of events
is the conventional method to search for global anisotropy
of cosmic ray arrival distribution.
Using this analysis, we have checked that large scale isotropy
predicted by our source models is consistent with isotropic source
distribution within a 90$\%$ confidence level with event number equal
to the current HiRes and AGASA observations.
On the other hand, arrival distributions observed by the HiRes and
AGASA do not also show any significant global anisotropy.
There is no discrepancy between the two experiments.
Thus we do not discuss large scale isotropy in the following.

Comparing figure \ref{image_s3} and \ref{image_s5}, we easily find
that arrival distribution expected for smaller source number density
exhibits stronger anisotropy on small angle scale.
This is simply because contributions from each source to the total
events are larger for smaller source number density.
Furthermore, the clustered events are aligned in the sky according to
the order of their energies, reflecting the direction of the GMF at
each direction.
This interesting feature becomes evident with increasing amount of the
event number.

%-------------------------------------------------------------------------
\subsection{Statistical significance of the small scale anisotropy}
\label{statistics_arrival}
%-------------------------------------------------------------------------

Next, we discuss the statistics on arrival distribution of UHECRs.
Figure \ref{2P_s3} and \ref{2P_s5} show the two point correlation
function expected in the case of source distributions of figure
\ref{image_s3} and \ref{image_s5}, respectively.
We calculate the two point correlation function for
the simulated events within only $0^\circ \le \delta \le 90^\circ$ in
accordance with the HiRes exposure.
For this reason, the event numbers written in figure \ref{2P_s3} and
\ref{2P_s5} differ from the ones in figure \ref{image_s3} and
\ref{image_s5}.
The error bars represent 1 $\sigma$ statistical fluctuations of our
numerical calculations due to the finite number of events.
The shaded regions represent 1 $\sigma$ fluctuations for uniform
source distribution.
The event selections are performed $1000$, $100$, and $30$ times for the
event number $150$, $1500$, and $5000$, respectively.
It is noted that we compare our numerical results with not the HiRes
data itself but the prediction of uniform source distribution.
We do not have to consider the non-uniform observation in the angular
sky.

\vspace{0.5cm}
%%%%%%%%%%%%%%%%%%%%%%%%%%%%%%%%%%%%%%%%%%%%%%%%%%%%%%%%%%%%%%%
\centerline{{\vbox{\epsfxsize=8.0cm\epsfbox{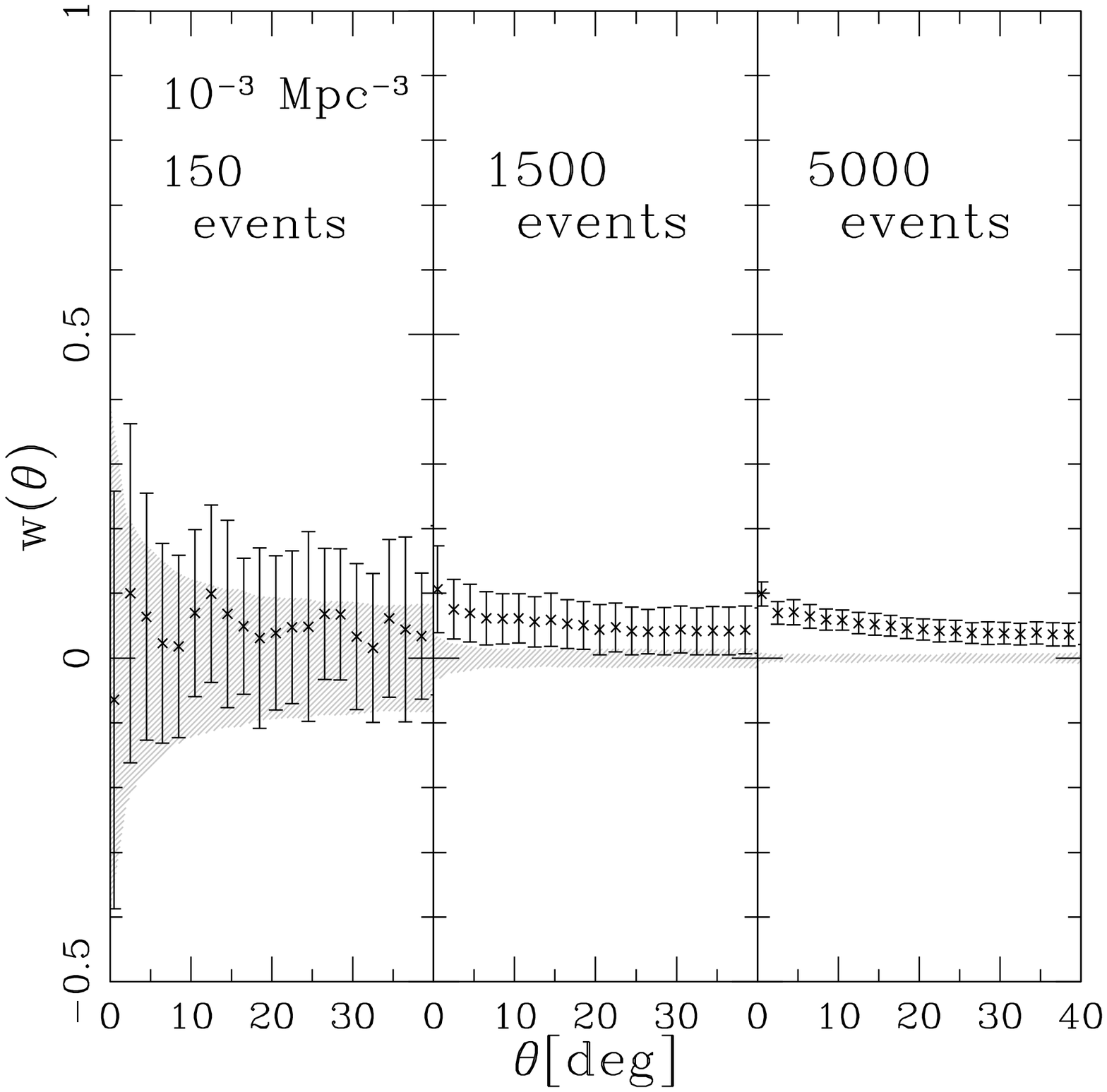}}}}
\figcaption{
Two point correlation function predicted by a realization of the
source selection for the number density $10^{-3}$ Mpc$^{-3}$.
We calculate the two point correlation function for
the simulated events within only $0^\circ \le \delta \le 90^\circ$ in
accordance with the HiRes exposure.
The error bars represent 1 $\sigma$ statistical fluctuations due to
the finite number of events for our numerical calculations.
The shaded regions represent 1 $\sigma$ fluctuations for uniform
source distribution.
The event selections are performed $1000$, $100$, and $30$ times for the
event number $150$, $1500$, and $5000$, respectively.
\label{2P_s3}}
\vspace{0.5cm}
%%%%%%%%%%%%%%%%%%%%%%%%%%%%%%%%%%%%%%%%%%%%%%%%%%%%%%%%%%%%%%%

As mentioned before, it is evident that smaller source number density
predicts stronger correlation between events at small angle scale.
The small scale anisotropy well reflects the number density of UHECR
sources.
It is already shown that the small scale anisotropy
above $E=4\times 10^{19}$ eV observed by the AGASA can be explained
with the source number density $\sim 10^{-5 \sim -6}$ Mpc $^{-3}$
assuming weak extragalactic magnetic field ($B \le 1$ nG, EGMF)
\citep*{yoshiguchi03a,yoshiguchi03c,blasi04}.
We thus discuss the statistical significance of the discrepancy
between the HiRes and AGASA by considering to what extent we can
determine the UHECR source number density from the current HiRes data.

Even if we specify the source number density, source distribution
itself can not be determined because of randomness when we select
galaxies from the ORS sample.
Therefore, we first evaluate the statistical significance of deviation
of the two point correlation function predicted by a source
distribution for a given number density from that expected for uniform
source distribution.
We then repeat such calculations for a number of realizations of the
source selection, and obtain the probability distribution of $\chi^2$
for a given source number density.

\vspace{0.5cm}
%%%%%%%%%%%%%%%%%%%%%%%%%%%%%%%%%%%%%%%%%%%%%%%%%%%%%%%%%%%%%%%
\centerline{{\vbox{\epsfxsize=8.0cm\epsfbox{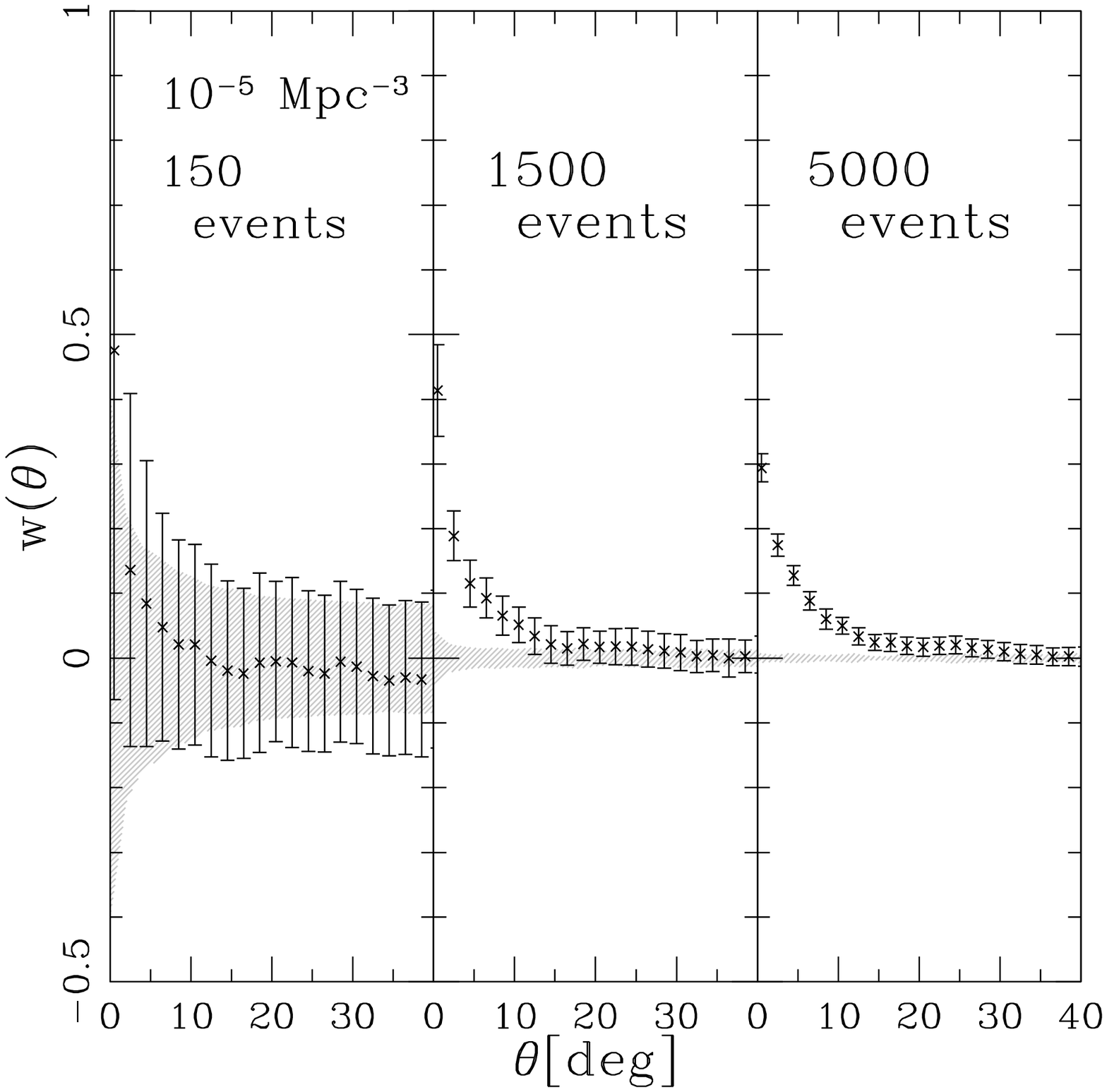}}}}
\figcaption{
Same as figure \ref{2P_s3}, but for the number density $10^{-5}$
Mpc$^{-3}$.
\label{2P_s5}}
\vspace{0.5cm}
%%%%%%%%%%%%%%%%%%%%%%%%%%%%%%%%%%%%%%%%%%%%%%%%%%%%%%%%%%%%%%%

\vspace{0.5cm}
%%%%%%%%%%%%%%%%%%%%%%%%%%%%%%%%%%%%%%%%%%%%%%%%%%%%%%%%%%%%%%%
\centerline{{\vbox{\epsfxsize=8.0cm\epsfbox{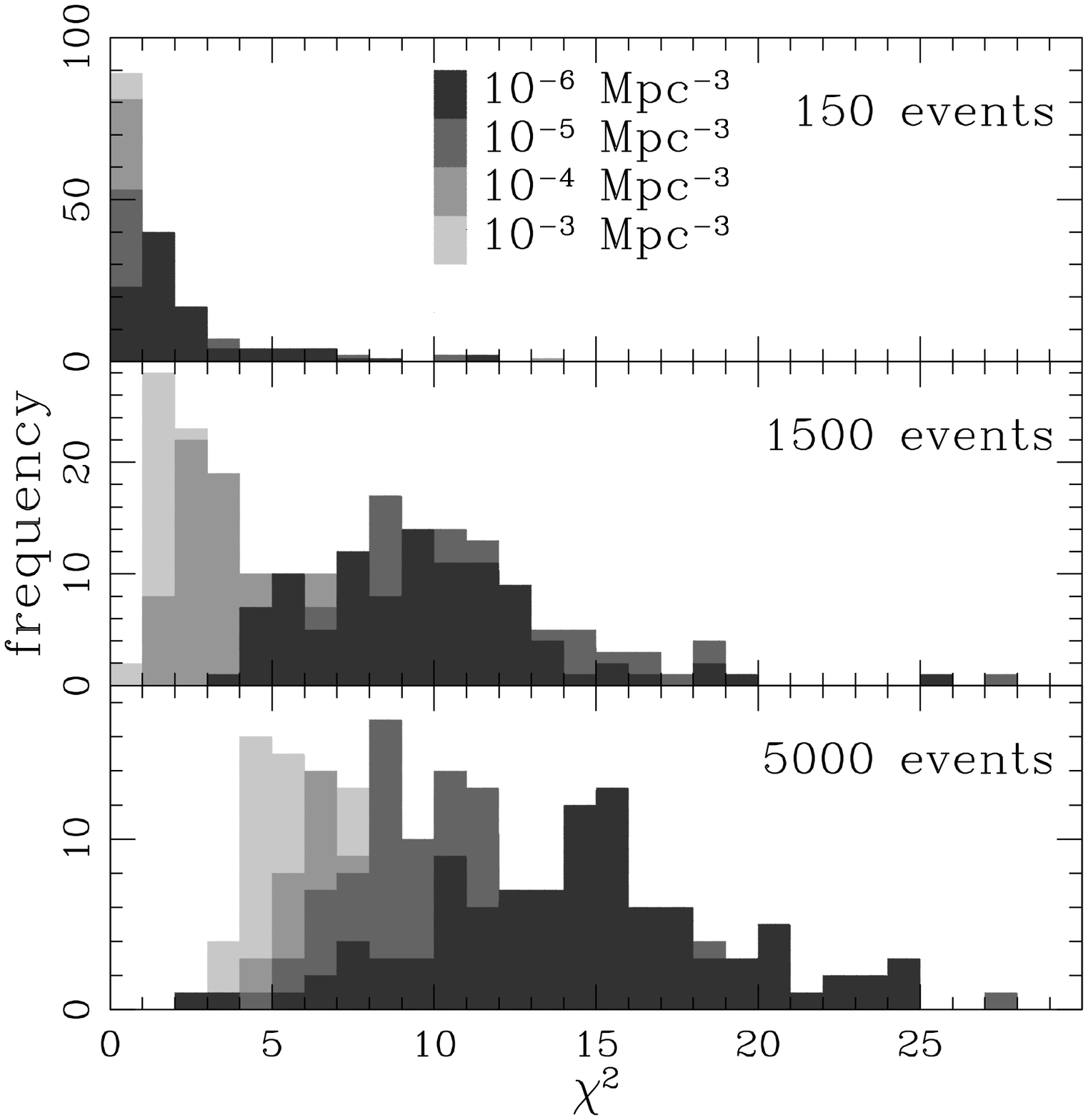}}}}
\figcaption{
Probability distribution of statistical significance of deviation of
our model prediction from that of isotropic sources
($\chi^2$) when the source selection from our ORS sample is performed
100 times.
The data of the two point correlation at the smallest angle are used
($N_{\rm bin} = 1$).
\label{histo}}
\vspace{0.5cm}
%%%%%%%%%%%%%%%%%%%%%%%%%%%%%%%%%%%%%%%%%%%%%%%%%%%%%%%%%%%%%%%

The results of the probability distribution of $\chi^2$ for $N_{\rm
bin} = 1$ and $5$ are shown as histograms in figure \ref{histo} and
\ref{histo5}, respectively.
The source selection is performed 100 times for all the source number
densities.
As we can see from the figures, differences between probability
distributions of $\chi^2$ for different source number densities are
clearer when $N_{\rm bin} = 1$.
This is related to the fact that the two point correlation function is
most sensitive to the source number density at the smallest angular
bin (see, figure \ref{2P_s3} and \ref{2P_s5}).
We thus focus our attention to the result for $N_{\rm bin} = 1$ in the
following.

\vspace{0.5cm}
%%%%%%%%%%%%%%%%%%%%%%%%%%%%%%%%%%%%%%%%%%%%%%%%%%%%%%%%%%%%%%%
\centerline{{\vbox{\epsfxsize=8.0cm\epsfbox{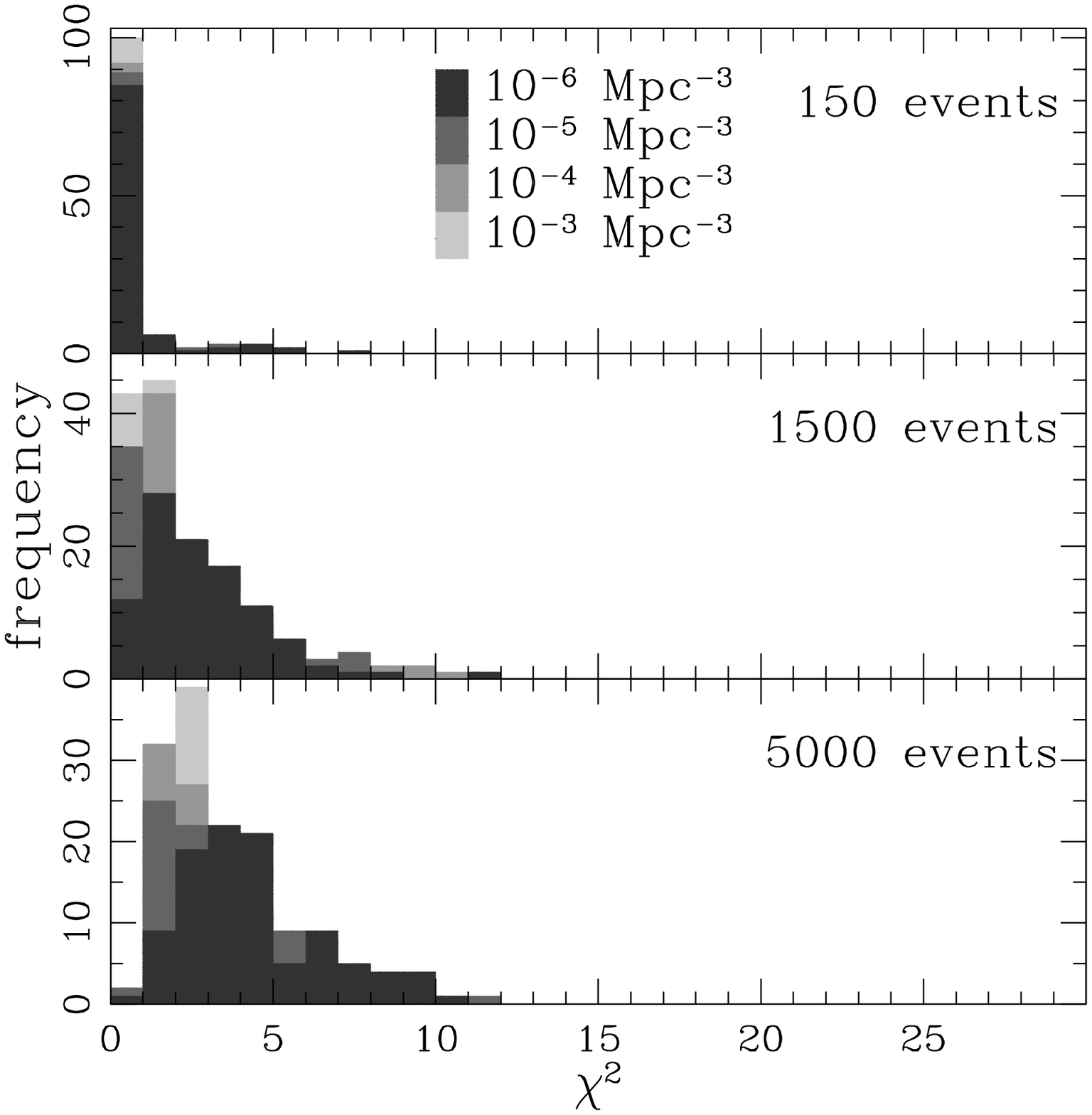}}}}
\figcaption{
Same as figure \ref{histo}, but for $N_{\rm bin} = 5$.
\label{histo5}}
\vspace{0.5cm}
%%%%%%%%%%%%%%%%%%%%%%%%%%%%%%%%%%%%%%%%%%%%%%%%%%%%%%%%%%%%%%%

As mentioned, we know that the small scale anisotropy observed
by the AGASA can be explained with the number density $\sim 10^{-6\sim
-5}$ Mpc$^{-3}$.
From figure \ref{histo}, it is found that the small scale
anisotropy predicted by this source number density is also consistent
with the prediction of isotropic source distribution when the event
number is equal to the HiRes data ($\sim 150$).
It is noted that the AGASA data have about 1000 events above $10^{19}$
eV.
The possibility that the number density is about $\sim 10^{-6\sim
-5}$ Mpc$^{-3}$ can not be ruled out, although the HiRes result seems
to be in agreement with the isotropic source distribution.
We thus conclude that the statement by the HiRes experiment that they
do not find small scale anisotropy in UHECR arrival distribution is
not statistically significant at present.

We now turn to the future prospects of determining the UHECR source
number density.
The probability distributions of $\chi^2$ with larger event number
than the current observations are also shown in figure \ref{histo}.
Of course, there is no observational data to be compared.
We thus compare the model predictions with the case of
isotropic source distribution.
The results should be interpreted as quantifying deviations of our
calculations from the prediction of isotropic sources in the unit of
statistical fluctuations.

As is evident from figure \ref{histo}, difference between the
probability distribution of $\chi^2$ for different source number
densities become clear when the number of data increases.
We will be able to determine the source number density by the
future observations of the small scale anisotropy.
In order to discuss to what accuracy we can know the source number
density, we need to calculate the probability distribution of $\chi^2$
more precisely with a large number of simulations of the source
selection.
This issue is beyond the scope of the present paper.
We will study it in a future publication.

%%%%%%%%%%%%%%%%%%%%%%%%%%%%%%%%%%%%%%%%%%%%%%%%%
\section{SUMMARY AND DISCUSSION} \label{summary}
%%%%%%%%%%%%%%%%%%%%%%%%%%%%%%%%%%%%%%%%%%%%%%%%%

In this paper, we considered the statistical significance of the
discrepancy between the HiRes and the AGASA experiment on the small
scale anisotropy in arrival distribution of UHECRs.
We calculated arrival distribution of UHECRs above $10^{19}$ eV
predicted by our source models constructed using the ORS galaxy
sample.
We applied the new method developed by us \citep{yoshiguchi03d} for
calculating arrival distribution with the modifications by the GMF.
The great advantage of this method is that it enables us to calculate
UHECR arrival distribution with lower energy ($\sim 10^{19}$ eV) than
previous studies by following only the trajectories actually reaching
the earth.

It has been realized that the small scale anisotropy in the UHECR
arrival distribution well reflects the number density of UHECR sources
\citep*{yoshiguchi03a,yoshiguchi03c,blasi04}.
And also, it is shown that the small scale anisotropy
above $E=4\times 10^{19}$ eV observed by the AGASA can be explained
with the source number density $\sim 10^{-5 \sim -6}$ assuming weak
extragalactic magnetic field ($B \le 1$ nG, EGMF) and that observed
UHE particles are protons
\citep*{yoshiguchi03a,yoshiguchi03c,blasi04}.
We thus took the source number density as a parameter of our source
model, and discussed the statistical significance of the small scale
anisotropy by considering to what extent we can determine the source
number density from the comparison of the model predictions with the
HiRes observation.

When we consider to what extent we can determine the source number
density of UHECRs, we have to quantify deviations of predictions
by our source models from the observation.
And also, even if we specify the source number density of our model,
source distribution itself can not be determined because of randomness
when we select galaxies from the ORS sample.
Therefore, we first evaluated the statistical significance of deviation
($\chi^2$) of the two point correlation function predicted by a source
distribution for a given number density from that expected for uniform
source distribution.
We then repeated such calculations for a number of realizations of the
source selection.
This gave the probability distributions of $\chi^2$ for various
number densities of UHECR sources.

The results for $N_{\rm bin} = 1$ and $5$ show that differences
between probability distributions of $\chi^2$ for different source
number densities are clearer when $N_{\rm bin} = 1$. 
This is related to the fact that the two point correlation function is
most sensitive to the source number density at the smallest angular
bin.
We thus focused our attention to the result of $N_{\rm bin} = 1$.
As mentioned above, we know that the small scale anisotropy observed
by the AGASA can be explained with the number density $\sim 10^{-6\sim
-5}$ Mpc$^{-3}$.
It is found that the small scale anisotropy predicted by this source
number density is also consistent with the prediction of isotropic
source distribution when the event number is equal to the HiRes data
($\sim 150$).
Note that the data number of AGASA is about $1000$ above $10^{19}$ eV.
The possibility that the number density is about $\sim 10^{-6\sim
-5}$ Mpc$^{-3}$ can not be ruled out, although the HiRes result seems
to be in agreement with the isotropic source distribution.
We thus concluded that the statement by the HiRes experiment that they
do not find small scale anisotropy in UHECR arrival distribution is
not statistically significant at present.

We also discussed the future prospects of determining the UHECR source
number density.
As is evident from figure \ref{histo}, difference between the
probability distribution of $\chi^2$ for different source number
densities become clear when the number of data increases.
We will be able to determine the source number density by the
future observations of the small scale anisotropy.
In order to discuss to what accuracy we can know the source number
density, we need to calculate the probability distribution of $\chi^2$
more precisely with a large number of simulations of the source
selection.
This issue deserves further investigation.

Finally, we mention the assumptions made in this paper.
We neglected the effects of the extragalactic magnetic field and the
random component of the GMF.
If we take these effect into account, the small scale anisotropy
obtained by numerical calculations become less obvious.
In this case, our statement that the small scale anisotropy for $\sim
10^{-5\sim -6}$ Mpc$^{-3}$ is also consistent with the prediction of
isotropic sources when the event number is equal to the HiRes data
($\sim 150$) remains to be valid.
Hence, the assumptions are not so important for our conclusion.

\vspace{0.3cm}
{\bf{Note added:}}
While we are finishing this study, a paper by the HiRes collaboration
\citep{abbasi04} appears on the net, where they present the results of
a search for small scale anisotropy in the observed arrival
distribution of UHECRs.
They find no small scale anisotropy in 271 events above $10^{19}$ eV,
which is slightly larger than the event number considered in the
present study (164).
However, this increase of the event number will not affect our
conclusion very much, as seen from figure~\ref{histo}.

%%%%%%%%%%%%%%%%%%%%%%%%%%%%%%%%%%%%%%%%%%%%%%%%%%%%%%%%%%%%%%%
%%%%%%%%%%%%%%%%%%%%%%%%%%%%%%%%%%%%%%%%%%%%%%%%%%%%%%%%%%%%%%%
\acknowledgments
%%%%%%%%%%%%%%%%%%%%%%%%%%%%%%%%%%%%%%%%%%%%%%%%%%%%%%%%%%%%%%%
%%%%%%%%%%%%%%%%%%%%%%%%%%%%%%%%%%%%%%%%%%%%%%%%%%%%%%%%%%%%%%%
The work of H.Y. is supported by Giants-in-Aid for JSPS Fellows.
The work of K.S. is supported by Giants-in-Aid for Scientific
Research provided by the Ministry of Education, Science and Culture
of Japan through Research Grant No.S14102004 and No.14079202.

%%%%%%%%%%%%%%%%%%%%%%%%%%%%%%%%%%%%%%%%%%%%%%%%%%%%%%%%%%%%%%%%%%%%%%%
%\clearpage
%%%%%%%%%%%%%%%%%%%%%%%%%%%%%%%%%%%%%%%%%%%%%%%%%%%%%%%%%%%%%%%%%%%%%%%

%%%%%%%%%%%%%%%%%%%%%%%%%%%%%%%%%%%%%%%%%%%%%%%%%%%%%%%%%%%%%%%%%%%%%%%

\end{document}